\documentstyle[12pt]{article}
\title{$K_S\rightarrow \gamma\gamma , K_L\rightarrow\pi^0\gamma\gamma$ and
Unitarity \thanks{Research supported by Schweizerischer Nationalfonds and
by the US National Science Foundation.}}
\author{Joachim Kambor and Barry R. Holstein\\[5mm]
Department of Physics and Astronomy\\
University of Massachusetts\\
Amherst, MA~01003}
\begin{document}
\begin{titlepage}
\maketitle
\begin{abstract}
Agreement between the experimental value $\Gamma (K_S\rightarrow \gamma\gamma)$
and the number predicted via a one-loop chiral perturbation theory
calculation has been cited as a success for the latter.  On the other hand
the one-loop prediction for the closely related process $K_L\rightarrow
\pi^0\gamma\gamma$ has been found to be a factor three below the
experimental value.  Using the inputs of unitarity and dispersion relations, we
demonstrate the importance of higher order loop effects
to both of these processes.
\end{abstract}
{\vfill
\begin{flushright}
UMHEP--397
\end{flushright}}
\end{titlepage}

\section{Introduction}
During the past decade, we have learned (at last) how to make rigorous contact
between experimental low energy hadronic physics and QCD which is presumed
to underlie such processes.  This contact is provided by chiral perturbation
theory (ChPT) \cite{GL}, which exploits the (broken) chiral invariance of the
light
quark component of the QCD Lagrangian and provides a representation for
interaction amplitudes as an expansion in
energy-momentum divided by the chiral scale
parameter $4\pi F_\pi \sim 1$ GeV \cite{BG}.  A review of ChPT techniques will
not be attempted here, but has been presented in a number of sources, wherein
it
is shown that, at least in the Goldstone boson sector, such a chiral
approach provides a remarkably successful and predictive representation
of a wide variety of experimental processes \cite{Hol}.

The chiral technology begins by writing down
an effective chiral Lagrangian, the simplest (two-derivative) form of
which is, in the Goldstone sector, \cite{GG}
\begin{equation}
{\cal L}_{\rm eff}^{(2)}={\bar{F}^2\over 4}{\rm Tr}D_\mu UD^\mu U^\dagger
+{\bar{F}^2\over 4}{\rm Tr}2B_0m(U+U^\dagger )+\cdots
\end{equation}
where
\begin{equation}
U=\exp\left({i\over \bar{F}}\sum_{j=1}^8\lambda_j\phi_j\right)
\end{equation}
is a nonlinear function of the pseudoscalar fields, $m=(m_u,m_d,m_s)_{\rm
diag}$
is the quark mass matrix,
\begin{equation}
2B_0={2m_K^2\over m_u+m_s}={2m_\pi^2\over m_u+m_d}={6m_\eta^2\over
m_u+m_d+4m_s}
\end{equation}
is a phenomenological constant, $D_\mu$ is the covariant derivative, and
$\bar{F}$ is the pion decay constant in the limit of chiral symmetry.  Although
these are only two of an infinite number of terms, already at this level there
exists predictive power--e.g., tree level evaluation of ${\cal L}^{(2)}$
yields the familiar Weinberg predictions (at ${\cal O}(p^2,m^2)$) for
S-wave $\pi -\pi$ scattering lengths \cite{Wei}
which are approximately borne out experimentally.  Loop diagrams required by
unitarity
produce terms of ${\cal O}(p^4,p^2m^2,m^4)$ and contain divergences.  However,
just as in QED such infinities can be absorbed into renormalizing
phenomenological
chiral couplings, and the most general "four-derivative" Lagrangian has been
given by Gasser and Leutwyler \cite{GL}
\begin{eqnarray}
{\cal L}^{(4)}_{\rm eff}&=&L_1({\rm Tr}D_\mu UD^\mu U^\dagger )^2
+L_2({\rm Tr} D_\mu UD_\nu U^\dagger )^2\nonumber\\
& &+L_3{\rm Tr}(D_\mu UD^\mu U^\dagger)^2 + \ldots
\end{eqnarray}
Here the bare $L_i$ coefficients are themselves unphysical and are related
to empirical quantities $L_i^r (\mu )$ measured at scale $\mu$ via
\begin{equation}
L_i^r(\mu )=L_i+{\Gamma_i\over 32\pi^2}\left({1\over \epsilon}+{\rm ln}
{4\pi\over \mu^2}+1-\gamma\right) ,
\end{equation}
where $\Gamma_i$ are constants defined in ref. \cite{GL} and
$\epsilon=4-d$ is the
usual parameter arising in dimensional regularization, with $d$ being
the number of dimensions.  Gasser and Leutwyler
have obtained empirical values for the phenomenological constants $L_1^r,
\ldots L_{10}^r$.

A wide range of electroweak and strong interactions of these Goldstone
bosons have been successfully treated within this formalism, but there
is at least one recurring problem---whenever the S-wave I=0 $\pi -\pi$
interaction is involved the simple
one loop predictions have in general been found to be wanting \cite{Mei}.
This is perhaps not
surprising, as the associated phase shift $\delta_0^0(s)$ passes through
$90^\circ$ somewhere in the vicinity of $\sqrt{s}\sim 700-900$ MeV,
which has sometimes been associated with the existence of a broad
resonance in this region.  Such resonant behavior can certainly not
be treated in any perturbative fashion and thus in this channel
the chiral expansion must break down well before this energy is reached.

There are a number of ways which have been used in order to avoid this
difficulty.  One is simply to confine predictions to a low enough energy
that one-loop corrections should be sufficient.  However, it is also
possible to treat such effects nonperturbatively either by inclusion
of S-wave I=0 ($\sigma$ or $\epsilon$) pole effects or by use of
dispersion relations which effectively treat the $\pi -\pi$ interaction
to all loop orders---but at the price of introduction of model dependence
\cite{GM}.
This cost is usually considered worth paying, however, as by inclusion
of such well-motivated model-dependence one can often
reliably extend the usual region
of validity of chiral predictions ($E<\sim400-500$ MeV) up to much
higher values ($E<\sim 1$ GeV).

A particularly useful example of this procedure has recently been
provided for the process $\gamma\gamma\rightarrow\pi^0\pi^0$.  Because only
neutral particles are involved there exists no tree level two-derivative or
four-derivative contribution to this reaction, which guarantees that the
one loop chiral perturbative prediction must itself be finite.  This has
been calculated as \cite{DHL}
\begin{equation}
{\rm Amp}(\gamma\gamma\rightarrow\pi^0\pi^0)=2e^2\epsilon^\mu_1
\epsilon^\nu_2 {s-m_\pi^2\over 16\pi^2F_\pi^2}\left({g_{\mu\nu}k_1\cdot k_2
-k_{2\mu} k_{1\nu} \over k_1\cdot k_2}\right)F(s,m_\pi^2).
\end{equation}
Here
\begin{eqnarray}
F(x,y)=1+{y\over x}\left[\ln \left({1+\beta (x)\over
1-\beta (x)}\right) -i\pi\right]^2=1+{y\over x}\ln^2\left({\beta(s)+1\over
\beta(s)-1}\right)\nonumber\\
\qquad{\rm with}\quad\beta (x)=\sqrt{x-4y\over x}.
\end{eqnarray}
However, the associated $\gamma\gamma\rightarrow\pi\pi$ cross section
is given by
\begin{equation}
\sigma (\gamma\gamma\rightarrow \pi^0\pi^0)={\alpha^2\over 256 \pi^3F_\pi^4}
{(s-m_\pi^2)^2\over s}\sqrt{1-{4m_\pi^2\over s}}|F(s,m_\pi^2)|^2
\end{equation}
and is found to bear little resemblance to its recently measured experimental
analog, as shown in Figure 1.

The solution to this problem has recently been explored by a number of authors
and has been found to be related to the inadequacy of the
one-loop approach in the I=0 S-wave
$\pi -\pi$ channel \cite{DH}.  The solution is most clearly presented in terms
of
a dispersion relation approach.
We assume, consistent with the
chiral expansion, that when we are in the near-threshold region the only
relevant higher order effects are in the helicity conserving S-wave amplitude,
which we write as
\begin{eqnarray}
\gamma\gamma\rightarrow\pi^{+}\pi^{-}:\qquad
f^{C}(s)&=&\frac{1}{3}\left[2f_{0}(s) +
f_{2}(s)\right]\nonumber\\
\gamma\gamma\rightarrow\pi^{0}\pi^{0}:\qquad
f^{N}(s)&=&\frac{2}{3}\left[f_{0}(s) -
f_{2}(s)\right],
\end{eqnarray}
where I = 0, 2 refers to the isospin of the final $\pi\pi$ state.  For neutral
pion production and working in the gauge wherein $\epsilon_{2}\cdot k_{2} =
\epsilon_{2} \cdot k_{1} = \epsilon_{1} \cdot k_{2} = \epsilon_{1} \cdot k_{1}
=
0$ the transition amplitude is
\begin{equation}
\gamma\gamma\rightarrow\pi^{0}\pi^{0}:\qquad  {\rm Amp} = 2ie^{2} \epsilon_{1}
\cdot
\epsilon_{2} f^{N}(s).
\end{equation}

In the charged pion case the Born and seagull contributions to this multipole
must also be included, so that the full amplitude becomes
\begin{eqnarray}
\gamma\gamma\rightarrow\pi^{+}\pi^{-}:\  {\rm Amp} =
2ie^{2}\left[\epsilon_{1}\cdot\epsilon_{2} a(s)
- \frac{\epsilon_{1}\cdot p_{+} \epsilon_{2}\cdot p_{-}}{p_{+} \cdot k_{1}} -
\frac{\epsilon_{1} \cdot p_{-}\epsilon_{2} \cdot p_{+}}{p_{+}
\cdot k_{2}}\right].
\end{eqnarray}
Here
\begin{equation}
a(s) = 1 + f^{C}(s) - f^{C}_{\rm Born}(s),
\end{equation}
where
\begin{equation}
f^{C}_{\rm Born}(s) = \frac{1-\beta^{2}(s)}{2\beta(s)} ln
\left(\frac{1+\beta(s)}{1-\beta(s)}\right) = f^{\rm Born}_{0}(s) = f^{\rm
Born}_{2}(s)
\end{equation}
is the Born approximation value for the helicity conserving S-wave multipole.
In the threshold region the phase of $f_{I}(s)$ is required by unitarity to
be equal to the corresponding $\pi\pi$ phase shift $\delta_{I}(s)$.  When
$s>16m^{2}_{\pi}$, inelastic reactions involving four pions are allowed.
However,
the inelasticity is small, being of order $E^{8}$ in the chiral expansion and
will be neglected.

The functions $f_{I}(s)$ are analytic functions of s except for cuts along
the positive and negative real axis. For positive $s$, the right hand cut
extends
from $4m^{2}_{\pi} < s < \infty$ and is due to the $s$ channel $\pi\pi$ state.
For
negative $s$, the left hand cut is due to $t,u$-channel intermediate states
such
as $\gamma\pi\rightarrow\pi\rightarrow\gamma\pi$ or
$\gamma\pi\rightarrow\rho\rightarrow\gamma\pi$, and extends from $-\infty<s<0$.
The single-channel final state unitarization problem has a simple solution in
terms of the Omn\`es function \cite{Om}
\begin{equation}
D_I^{-1}(s)=\exp\left({s\over \pi}\int^\infty_{4m_\pi^2} {ds'\over s'}
{\delta_0^I(s')\over s'-s-i\epsilon}\right)
\end{equation}
---the result must have the form
\begin{equation}
f_I(s)=g_I(s)D_I^{-1}(s)
\end{equation}
where $g_{I}(s)$ is an analytic function with no cuts along the positive real
axis.  Morgan and Pennington consider a function $p_{I}(s)$ which has the same
left hand singularity structure as $f_{I}(s)$, but which is real for $s>0$.
They
then write a twice subtracted dispersion relation for the difference
$(f_{I}(s)-p_{I}(s))D_{I}(s)$, with the result \cite{MP}
\begin{equation}
f_I(s)=D_I^{-1}(s)\left[ p_I(s)D_I(s)+(c_I+sd_I)-{s^2\over \pi}\int^\infty
_{4m_\pi^2}{ds'\over {s'}^2}{p_I(s'){\rm Im}D_I(s')\over
s'-s-i\epsilon}\right],
\end{equation}
where $c_{I},d_{I}$ are subtraction constants.  Picking $p_I(s)$ to be given
by its Born value and matching onto the known form of the low energy amplitude
required by chiral symmetry determines these constants unambiguously to be
\begin{equation}
c_I=0,\quad d_I={2\over F_\pi^2}(L_9^r+L_{10}^r)+{1\over 384\pi^2}\times
\left\{\begin{array}{ll}
             -1 & I=0 \\
             +2 & I=2
         \end{array}\right .
\end{equation}

One can now address the origin of the large corrections found in
the $\gamma\gamma\rightarrow\pi^{0}\pi^{0}$ amplitude. Do they arise simply
from
the unitarization of the amplitude ({\it i.e.} $D_{I}(s)\neq 1$) or are new
inputs
needed in the amplitude?
It turns out that the rescattering physics in $D^{-1}_{I}(s)$ is most
important,
and that the main corrections are due to well-known ingredients.  In our
subsequent discussion, we will use a full phenomenological treatment but
it is useful here to
explore the case with a simple analytic form for
$D^{-1}_{I}(s)$.  The condition $Im D_{I}(s)=-\beta (s)t^{\rm CA}_{I}(s)$
defines the
[0,1] Pad\'{e} approximation for the Omn\`es function \cite{Tru1}, {\it i.e.}
\begin{eqnarray}
D_I^{-1}(s)&=&{1\over 1-k_Is+t_I^{\rm CA}(s)(h(s)-h(0))}\nonumber\\
{\rm with} \qquad h(s)&=&{\beta (s)\over \pi}ln\left( {\beta (s) +1\over
\beta(s)
-1}\right), \qquad h(0)={2\over \pi}
\label{DPade}
\end{eqnarray}
and allows one an approximate but  simple analytic representation for the
$\gamma\gamma\rightarrow\pi\pi$ amplitude.  The constant $k_{0} \cong
\frac{1}{25m_{\pi}^{2}}$ is
chosen to match the small $s$ behavior of the experimental $D^{-1}_{0}(s)$
function, and $k_{2}\cong -\frac{1}{30m_{\pi}^{2}}$ is chosen from a fit to $I
=
2$ $\pi\pi$ scattering.  The resulting form for the
$\gamma\gamma\rightarrow\pi^{0}\pi^{0}$
amplitude is
\begin{eqnarray}
f^N(s)&=&-{1\over 48\pi^2F_\pi^2}F(s,m_\pi^2)\nonumber\\
&\times &\left[(2s-m_\pi^2)D_0^{-1}(s)+
(s-2m_\pi^2)D_2^{-1}(s)\right]\nonumber\\
&+&{4\over 3F_\pi^2}(L_9^r+L_{10}^r)s
(D_0^{-1}(s)-D_2^{-1}(s))
\end{eqnarray}
which, when the Pad\'{e} forms of $D^{-1}_I(s)$ are used, provides a
consistent analytic solution to the dispersion relation while also displaying
the
correct chiral properties to ${\cal O}(s)$.  In Figure 1, we plot the resulting
cross section, in
comparison with the data and the lowest order result. It can be seen that the
Omn\`es functions produce a substantial modification even near threshold. Of
these,
the most important is $D^{-1}_{0}(s)$ which reflects the strong attractive
$\pi\pi$ scattering in the $I=0, J=0$ channel \cite{Mei}.

A much more satisfactory fit is found by use of an Omn\`es function
$D_0^{-1}(s)$
determined via the use of {\it experimental} values of the pion-pion phase
shifts \cite{DGL} as
well as including contributions to the left hand cut due to $A1,\rho ,\omega$
exchange diagrams, which leads to the very good fit given in Figure 1. Details
of this
calculation can be found in ref. \cite{DH}.

The lesson to be learned from this example is the importance of inclusion of
I=0 S-wave $\pi -\pi$ rescattering corrections especially in processes which
have no counterterm contributions and are generated from a simple one-loop
chiral
calculation.  We shall see in the next sections two additional examples of this
type,
when we extend our formalism to cover the nonleptonic weak-radiative decays
$K_S\rightarrow \gamma\gamma$ and $K_L\rightarrow \pi^0\gamma\gamma$.

\section{$K_S\rightarrow \gamma\gamma$}

A good deal of work has been done extending the chiral formalism into the
regime
of nonleptonic weak processes.  To two-derivative order the form of the
effective SU(3) octet chiral Lagrangian is unique
\begin{equation}
{\cal L}_{\Delta S=1}=F_\pi^4G_8{\rm Tr}\left(\lambda_6D_\mu UD^\mu
U^\dagger\right)
\end{equation}
where $G_8$ is a constant whose value can be determined empirically.
The corresponding four-derivative weak effective Lagrangian
has also been written down and
involves thirty-seven
additional terms \cite{KMW}, whose explicit form will not be needed here, since
as in the
case of $\gamma\gamma\rightarrow\pi^0\pi^0$ there is no tree level
contributions to the weak-radiative decay process $K_S\rightarrow
\gamma\gamma$.   Nevertheless there does exist a finite one-loop piece.
Defining
\cite{Goi}
\begin{equation}
{\rm Amp}(K_S\rightarrow \gamma\gamma)=\epsilon_1^\mu\epsilon_2^\nu
\left( {-q_1\cdot q_2
g_{\mu\nu}+q_{1\nu}q_{2\mu}\over
q_1\cdot q_2}\right) \alpha F_\pi B(m_K^2 )
\end{equation}
the one-loop chiral prediction is found to be
\begin{equation}
B(s)=G_8^{CA}\left({1\over \pi}(m_\pi^2-s)F(s,m_\pi^2)-[m_\pi^2\rightarrow
m_K^2]\right).
\label{ChPT:B}
\end{equation}

With the value
\begin{equation}
G_8^{CA}\approx 9.1\times 10^{-6} GeV^{-2}
\end{equation}
determined from the tree level prediction for $K_S\rightarrow\pi\pi$
\begin{equation}
{\rm Amp}^8(K_S\rightarrow \pi^+\pi^-)\equiv 2F_\pi A_0^{CA}(m_K^2)=
2F_\pi G_8^{CA}(m_K^2-m_\pi^2)
\end{equation}
we find the one-loop chiral prediction to be in good agreement with the
recently determined experimental number
\begin{eqnarray}
{\rm B.R.}(K_S\rightarrow \gamma\gamma)_{\rm ChPT}&=&2.0\times 10^{-6}
\qquad\quad {\rm vs.}\nonumber \\
{\rm B.R.}(K_S\rightarrow\gamma\gamma )_{exp}&=&
(2.4\pm 1.2)\times 10^{-6}.
\end{eqnarray}

An alternative derivation of this result is provided by the
use of a dispersion relation.  We note that the function B(s) has a cut
along the line element $4m_\pi^2<s<\infty$ and by unitarity has the
imaginary part \cite{Gob}
\begin{equation}
{\rm Im}B(s)=\theta (s-4m_\pi^2)\beta(s) A_0(s) f_0^*(s),
\label{ImB}
\end{equation}
where $A_0(s)$ is the amplitude for the weak decay
$K_S\rightarrow\pi\pi (I=0)$ and $f_0(s)$ is the amplitude for
S-wave radiative pion annihilation $\pi\pi (I=0)\rightarrow \gamma\gamma$
discussed in the previous section.
Note that since we are using an SU(3) octet assumption for the weak transition,
both
di-pions are required to be in an isoscalar configuration.  To lowest
order in chiral perturbation theory we have
\begin{eqnarray}
A(s)&\approx& A_0^{\rm CA}(s)=G_8^{CA}(s-m_\pi^2)\nonumber\\
f_0(s)&\approx &f_{\rm Born}(s)={1-\beta^2(s)\over 2\beta (s)}\ln
\left({1+\beta (s)
\over 1-\beta (s)}\right).
\label{loword}
\end{eqnarray}
We now write a doubly subtracted dispersion relation for the function $B(s)$
using as subtraction constants the requirements that the amplitude vanish
at both $s=0$ and $s=m_\pi^2$, as given in the lowest order chiral analysis,
yielding
\begin{equation}
B(s)={s(s-m_\pi^2)\over \pi}\int_{4m_\pi^2}^\infty {{\rm Im} B(s')ds'
\over s'(s'-m_\pi^2)(s'-s-i\epsilon)}.
\label{Disp:B}
\end{equation}
If the lowest order chiral values---Eq. \ref{loword}---are used to determine
Im $B(s')$ then the integration can be performed analytically to yield
\begin{equation}
B(s)={1\over \pi}(m_\pi^2-s)F(s,m_\pi^2)G_8^{CA},
\end{equation}
which is precisely the (pion contribution to the) one-loop chiral result Eq.
\ref{ChPT:B}.

At one level then the dispersive technique represents merely an
alternative (and simpler!) way by which to perform the one-loop calculation.
However,
at a deeper level the dispersion relation provides a means to
undertake a much more complete calculation of the radiative decay
process by using for Im $B(s')$ not just the lowest order chiral
forms for these amplitudes but instead values which have more
experimental validity.  Of course, the subtraction constants must be fixed
by some other means --- we assume that chiral perturbation theory to one
loop is accurate enough to describe the amplitude at very low energies.

The $K_S\rightarrow \pi\pi$ decay amplitude $A_0(s)$ itself is
an analytic function with a discontinuity along a cut $4m_\pi^2<s<\infty$
given in terms of the S-wave I=0 $\pi -\pi$ scattering phase shift
\begin{equation}
{\rm Im}A_0(s)=\theta (s-4m_\pi^2)e^{-i\delta_0^0(s)}\sin \delta_0^0(s)A_0(s).
\end{equation}
The general solution of such an equation is given
\begin{equation}
A_0(s)=P(s)D_0^{-1}(s),
\label{A0Omn}
\end{equation}
where $P(s)$ is an arbitrary polynomial and $D_0^{-1}(s)$
is the Omn\`es function discussed in section 1.  In order to determine the
polynomial $P(s)$ we demand that the full $K_S\rightarrow\pi\pi$ amplitude
given in Eq. \ref{A0Omn} match the simple chiral form Eq. \ref{loword}
in the absence of
rescattering ({\it i.e.} when $D_0^{-1}(s)=1$).  We have then
\begin{equation}
A_0(s)=G_8(s-m_\pi^2)D_0^{-1}(s)
\label{A0match}
\end{equation}
while the corresponding solution for $f_0(s)$ was derived in ref. \cite{DH}
and has been outlined in section 1 of this paper.\footnote {It is important
to note here that when the solution Eq. \ref{A0match} for $A_0(s)$ is used,
the value
$G_8=G_8^{CA}/|D_0^{-1}(s=m_K^2)|\approx 6.1\times 10^{-6}GeV^{-2}$
must be employed in order that the proper normalization
to the $K_S\rightarrow\pi\pi$ decay rate be preserved.}
Substitution into
Eq. \ref{ImB} and numerical evaluation of the dispersive integral then provides
a prediction for the $K_S\rightarrow\gamma\gamma$ amplitude which is
much more complete and founded in experiment than is its simple one-loop
chiral analog.  The result of the numerical integration is shown in
Table 1, where values are given for the radiative decay
amplitude $B(m_K^2)$/branching ratio for four scenarios---a) simple one-loop
chiral
perturbation theory; b) use of the full unitarized values and
with the Pad\'{e} form for the Omn\`es function; c) use of the full
unitarized values and with a numerical representation of the Omn\`es function
based on experiment and given by Gasser \cite{DGL,GM};
d) same as c) but with vector and axial-vector
exchange contributions included also in $f_0(s)$, which provides
the best fit to the $\gamma\gamma\rightarrow\pi\pi$ system.

\begin{table}
\begin{center}
\begin{tabular}{|l|l|l|l|}
\hline
{\em Input}     &  Im $B$ $[10^{-6}]$ & Re $B$ $[10^{-6}]$ & $BR$ $[10^{-6}]$\\
\hline
ChPT  &  0.77  &  $-0.44$  &  2.0 \\
\hline
Pad\'{e}  &  0.47  &  $-0.83$  &  2.3 \\
\hline
Gasser & 0.49  & $-0.73$  &  2.0 \\
\hline
Gasser+V,A  & 0.46 & $-0.84$  &  2.3 \\
\hline
\end{tabular}
\caption{Shown are the amplitudes and predicted branching ratios
for the process $K_S\rightarrow\gamma\gamma$ using various theoretical
inputs to the dispersion relation Eq. 28 as described in the text.}
\end{center}
\end{table}

Examination of this table shows that the full unitarization procedure
makes substantial changes in the predicted form of the decay amplitude
with the importance of the real/imaginary pieces being interchanged
for the ChPT and full unitarized calculations respectively.
This is clearly seen in Figures 2 where we show the
very different shapes for Im,Re $B(s)$ for these cases.
Despite these differences the predicted branching ratio is remarkably
robust, changing only slightly among the very different scenarios.
We conclude then that the agreement between the experimental
$K_S\rightarrow \gamma\gamma$ rate and that predicted via one-loop
ChPT should {\it not} be considered as a success for the latter,
since we have produced nearly identical predictions for the branching ratio
from a very different set of assumptions concerning the input parameters.
Truong has reached a similar conclusion using a different parametrization
of the Omn\`es function and an approximated $\gamma\gamma \rightarrow \pi\pi$
amplitude \cite{Tru3}.

Therefore, one must conclude that in order to distinguish between
the various decay mechanisms one must examine a process which offers the
chance for a rather richer experimental confrontation---the
related nonleptonic radiative decay $K_L\rightarrow\pi^0\gamma\gamma$.

\setcounter{section}{2}
\setcounter{equation}{32}

\section{$K_L\rightarrow\pi^0\gamma\gamma$}

There has been considerable recent interest in the nonleptonic-radiative
mode $K_L\rightarrow\pi^0\gamma\gamma$.  This began when one-loop chiral
perturbation theory was used to generate a supposedly reliable prediction
\cite{EPR}
\begin{equation}
{\rm Amp}(K_L\rightarrow\pi^0\gamma\gamma )=\epsilon_1^\mu\epsilon_2^\nu
\left({-g_{\mu\nu}k_1\cdot k_2 +k_2^\mu k_1^\nu\over k_1\cdot k_2}\right)
\alpha C(s)
\label{AmpKpigg}
\end{equation}
where
\begin{equation}
C(s)={1\over 2\pi}G_8^{CA}\left[F(s,m_\pi^2)(m_\pi^2-s)
-F(s,m_K^2)(m_K^2+m_\pi^2-s)\right].
\label{ChPT:C}
\end{equation}

Since a three-body final state is involved, what emerges is a prediction
for both the overall branching ratio
\begin{equation}
B.R.(K_L\rightarrow\pi^0\gamma\gamma )= 0.68\times 10^{-6}
\label{BRChPT}
\end{equation}
in addition to the shape of the decay spectrum, as shown in
Figure 3.
This distinctive shape is in marked contrast to that arising from a
simple $\eta ,\eta'$ pole model, which gives support at lower values
of $s_{\gamma\gamma}$ \cite{Seh}. When experimental numbers were provided, the
shape was found to be in good agreement with the ChPT prediction.
However, the measured rate was nearly a factor of three larger than
given in Eq. \ref{BRChPT}
\begin{equation}
B.R.(K_L\rightarrow\pi^0\gamma\gamma)= \left\{\begin{array}{ll}
(1.7\pm 0.3)\times 10^{-6} & {\rm NA31} \cite{Bar}\\
(1.86\pm 0.6)\times 10^{-6} & {\rm FNAL} \cite{Pap}.
\end{array}\right.
\end{equation}
Since this finding a number of authors have examined this problem. The
inclusion of the $\Delta I=3/2$ weak interaction results in a minor effect,
as expected \cite{CDM}. A dispersive analysis including unitarity corrections
of ${\cal O}(p^6)$ due to the $\pi^+\pi^-$ intermediate state has also been
presented recently \cite{CEP}. In this approach, the experimental results for
the branching ratio as well as
the spectrum in the invariant mass of the two photons can be reproduced if
a somewhat sizeable contribution of vector meson exchange to the counterterm
lagrangian
of ${\cal O}(p^6)$ is assumed. A similar result has been obtained in
ref. \cite{HS},
however without taking into account unitarity corrections of ${\cal O}(p^6)$.

Here we wish to examine the contribution of higher order diagrams to the
decay process, by generalizing the dispersive approach which was applied in the
previous section to $K_S\rightarrow\gamma\gamma$. By definition only the
amplitude proportional to the Lorentz structure in Eq. \ref{AmpKpigg} enters
the calculation. However, for this amplitude, unitarity corrections due to
$\pi\pi$ intermediate states will be treated to all orders in the chiral
expansion.  We begin by rederiving the
one-loop ChPT result in this fashion.  The amplitude $C(s)$ possesses a
cut along the real axis from $4m_\pi^2<s<\infty$ with a discontinuity
determined
via unitarity, which takes the form \cite{KR}
\begin{equation}
{\rm Im}C(s)=\theta (s-4m_\pi^2){1\over 2}\beta(s) A_{+-0}^{CA}(s) f^{C*}(s)
\end{equation}
where the lowest order ChPT amplitudes for $\gamma\gamma\rightarrow
\pi^+\pi^-$
and $K_L\rightarrow \pi^+\pi^-\pi^0$ are given respectively by
\begin{eqnarray}
f^C(s)&=&f^{\rm Born}(s), \nonumber \\
{\rm Amp}(K_L\rightarrow \pi^+\pi^-\pi^0)&\equiv &A_{+-0}^{CA}(s)=G_8^{CA}
(s-m_\pi^2).
\label{K3piCA}
\end{eqnarray}

As for $K_S\rightarrow \gamma\gamma$ we now write a twice subtracted
dispersion relation for the function $C(s)$, the subtraction constants
being specified by the one-loop ChPT requirement that $C(s)$ vanishes at
both $s=0$ and $s=m_\pi^2$:
\begin{equation}
C(s)={s(s-m_\pi^2)\over \pi}\int_{4 m_\pi^2}^\infty {{\rm Im}C(s') ds'\over
s'(s'-m_\pi^2)(s'-s-i\epsilon) }.
\label{disp:C}
\end{equation}
Using the lowest order chiral expression to determine Im $C(s)$ the integration
can again be performed analytically yielding
\begin{equation}
C(s)={1\over 2\pi}G_8^{CA}F(s,m_\pi^2)(s-m_\pi^2)
\end{equation}
which reproduces exactly the (pion loop contribution to
the) one-loop ChPT result, {\it i.e.} the first term in Eq. \ref{ChPT:C}.

To provide an improved estimate, we use the dispersion relation Eq.
\ref{disp:C}, but with a more complete
representation of Im $C(s)$ than just the lowest order ChPT expression (the
subtraction constants are still taken from one-loop ChPT). In addition to the
$\pi^+\pi^-$ intermediate state considered so far we include the discontinuity
from the $\pi^0\pi^0$ intermediate state in the s-channel to the unitarity
relation
\footnote{This intermediate state does not contribute to Im $C$ in the chiral
one-loop analysis since $f^N$ vanishes at order $p^2$. However the two neutral
pions also have a $I=0$ component, therefore final state interactions in this
channel are expected to give large corrections.}
\begin{equation}
{\rm Im}C(s)=\theta (s-4m_\pi^2){1\over 2}\beta (s)
\left[ A_{+-0}^S(s) f^{C*}(s)+{1\over 2!}A_{000}^S(s) f^{N*}(s) \right] ,
\label{ImCimproved}
\end{equation}
where the superscript S indicates the S-wave component of these
amplitudes.
Other intermediate states open up at much higher thresholds and are
suppressed in the twice subtracted dispersion relation.

To evaluate the improved imaginary part of the function $C(s)$ in
Eq. \ref{ImCimproved}
we shall employ dispersion relations to calculate the $\gamma\gamma
\rightarrow\pi\pi$ scattering amplitudes $f^C$, $f^N$ and the $K\rightarrow
3 \pi$ decay amplitudes $A_{+-0}$, $A_{000}$. The approach to $f^C$, $f^N$
has been reviewed in section 1. We shall use the explicit results given
there.  As for the $K\rightarrow 3\pi$ amplitude we shall use an
approximate solution to twice subtracted Khuri-Treiman equations \cite{KT}.
Define the $\Delta I={1\over 2}$ $K\rightarrow 3\pi$ decay amplitude as
\begin{eqnarray}
{\rm Amp}(K^0_L\rightarrow \pi^a\pi^b\pi^c)= \delta^{ab}\delta^{c3}F(s_a,
s_b,s_c)+{\rm permutations}
\end{eqnarray}
where $s_i=(k-q_i)^2$.  Also note that $F(s_a,s_b,s_c)$ must be symmetric in
its first two arguments according to Bose statistics. The decay modes
relevant to our calculation are then
\begin{eqnarray}
{\rm Amp}(K^0_L\rightarrow \pi^+\pi^-\pi^0)\equiv A_{+-0}&=&F(s_a,s_b,s_c),
\nonumber\\
{\rm Amp}(K^0_L\rightarrow 3 \pi^0)\equiv A_{000}&=&F(s_a,s_b,s_c)+
{\rm permutations}.
\end{eqnarray}
The discontinuity of $F(s_a,s_b,s_c)$ is provided by unitarity. Thus for
$K\rightarrow 3\pi$ the unitarity condition reads in general
\begin{eqnarray}
\lefteqn{
{\rm Im}<\pi^a\pi^b\pi^c|{\cal H}_w^{1\over 2}|K_L^0>=} \nonumber\\
& &{1\over 3!}\sum_{def}
\int {d^3p_d\over (2\pi)^32E_d}\cdots{d^3p_f\over (2\pi )^32E_f}
(2\pi )^4\delta (k-p_d-p_e-p_f)\nonumber\\
& &\times <\pi^a\pi^b\pi^c|\pi^d\pi^e\pi^f>^*<\pi^d\pi^e\pi^f|{\cal H}_w
^{1\over 2}|K_L^0>.
\end{eqnarray}
However, this equation is not amenable to an exact solution, so various
approximations are necessary.  We begin by approximating the $3\pi$
scattering amplitude by a sum of two-pion, one spectator amplitudes---
\begin{equation}
<\pi^a\pi^b\pi^c|\pi^d\pi^e\pi^f>\sim \delta^{cf}<\pi^a\pi^b|\pi^d\pi^e>
+{\rm permutations}.
\end{equation}
Also, since any two-pion reaction is at low energy we include only S-wave
 scattering terms
\begin{eqnarray}
<\pi^a\pi^b|\pi^d\pi^e>\approx \delta^{ab}\delta^{de}{1\over 3}
(A^{(0)}(s)-A^{(2)}(s))\nonumber\\
+{1\over 2}(\delta^{ad}\delta^{be}+\delta^{ae}\delta^{bd})A^{(2)}(s)
\end{eqnarray}
where
\begin{equation}
A^{(I)}(s)=e^{i\delta^I_0(s)}\sin\delta_0^I(s).
\end{equation}
The unitarity relation then reduces to the simpler form
\begin{eqnarray}
{\rm Im}F(s,t,u)&=&  \int {d\Omega\over 4\pi} [
A^{(0)*}(s)F(s,t,u) \nonumber\\
&+&{1\over 3}(A^{(0)*}(s)-A^{(2)*}(s))(F(u,s,t)+F(t,u,s)) \nonumber \\
&+&{1\over 2} A^{(2)*}(t)(F(s,t,u)+F(u,s,t)) \nonumber \\
&+&{1\over 2} A^{(2)*}(u)(F(s,t,u)+F(t,u,s)) ].
\end{eqnarray}
The Khuri-Treiman equations are obtained by the ansatz $F(s,t,u)=U(s)+V(t)+
V(u)$. We use a twice subtracted form with linear subtraction polynomials
\begin{eqnarray}
U_0(s)&=&a_U+b_Us, \nonumber\\
V_0(s)&\equiv& 0.
\end{eqnarray}
The resulting system of equations
has a simple approximate solution provided that we ignore the generally small
$I=2$ scattering term with respect to its much larger $I=0$ counterpart:
\begin{eqnarray}
U(s)&=&U_0(s)+\hat U(s_1)\left({D_0(s_1)\over D_0(s)}-1\right)
{s-s_2\over s_1-s_2}\nonumber\\
    & &\qquad\quad  +\hat U(s_2)\left({D_0(s_2)\over D_0(s)}-1\right)
{s-s_1\over s_2-s_1} +\Phi_2, \nonumber\\
V(s)&=&0.
\label{NSsolution}
\end{eqnarray}
where $s_1$,$s_2$ are the subtraction points, $\Phi_2$ is a correction which
can be shown to be small, $D^{-1}_0$ is the Omn\`es function and \cite{NS}
\begin{equation}
\hat U(s)=U_0(s)+{2\over 3}\int {{d \Omega\over 4\pi} U_0(t(s,\cos \theta))}.
\end{equation}

We still need to specify the subtraction polynomial $U_0(s)$. Since we want
to input accurately the information provided by unitarity, we would like to use
as much as possible the experimental information on $K_L\rightarrow 3\pi$
decays. The $K\rightarrow 3\pi$ amplitude may be expanded around the center
of the Dalitz plot as (neglecting $\Delta I=3/2$ contributions)
\begin{equation}
A_{+-0}=\alpha_1-\beta_1 Y+(\zeta_1+\xi_1) Y^2+{1\over 3}(\zeta_1-\xi_1) X^2
\label{DPexpansion}
\end{equation}
with
\begin{equation}
Y={s_3-s_0\over m_\pi^2}, \qquad X={s_2-s_1\over m_\pi^2}.
\end{equation}
Assuming the coefficients $\alpha_1, \beta_1, \ldots$ are real, experiment
fixes $\alpha_1,\beta_1$ at the values given in column one of Table 2.  As
is well known, using the current algebra expression
$U_0(s)=G_8^{CA}(s-m_\pi^2)$
as a subtraction polynomial gives values for $\alpha_1$ and $\beta_1$ which are
too small.  On the other hand,  ChPT to one-loop can fit the
$K\rightarrow 3\pi$ data by adjusting the counterterm coupling constants of
order $p^4$. In the language of dispersion relations, this means that beyond
leading order ChPT not only final state interactions contribute to coefficients
$\alpha_1$, $\beta_1$ but also the subtraction polynomial $U_0$ is
subject to corrections. This might also include contributions from higher
resonance exchange.

\begin{table}
\begin{center}
\begin{tabular}{|l|r|r|r|}
\hline
[$10^{-8}$] & {\em exp.} & {\em Fit (Pad\'{e})} & {\em Fit (Gasser)} \\
\hline
$\alpha_1$  &  \ $91.7\pm 0.3$  &  \ $89.6+i19.6$  & \ $89.9 +i18.2$ \\
\hline
$\beta_1$  &  $-25.7 \pm 0.3$  &  $-21.8 -i13.6$  &  $-22.6-i12.2$\\
\hline
\end{tabular}
\caption{Shown are values for the $\Delta I={1\over 2}$ component
of the $K_L\rightarrow 3\pi$ decay amplitude with final state interactions
generated by the Pad\'{e} and Gasser forms for $D_0^{-1}$ respectively.}
\end{center}
\end{table}

In the same spirit we adopt the following phenomenological
procedure. The $K\rightarrow 3\pi$ decay amplitude is calculated according
to the approximate solution of Khuri-Treiman equations,
Eq. \ref{NSsolution}. The
subtraction constants $a_U$, $b_U$ of $U_0$ are treated as free
adjusted such that the experimentally found $\alpha_1$, $\beta_1$ are
reproduced. The values needed for the case of the Pad\'{e} form of $D_0^{-1}$
are
\begin{equation}
a_U^{\rm fit}=-2.3\cdot 10^{-7},
\qquad b_U^{\rm fit}=8.6\cdot 10^{-6} {\rm GeV}^{-2}
\end{equation}
to be compared with the current algebra expressions
$a_U^{\rm CA}=-G_8 m_\pi^2=-1.8\cdot 10^{-7}$, $b_U^{\rm CA}=G_8=9.1\cdot
10^{-6} {\rm GeV}^{-2}$.
\footnote{In a similar approach, ref. \cite{Tru2}, Truong obtains a reasonable
representation of the $K\rightarrow 3\pi$ decay amplitude using current
algebra constraints for subtraction constants and appending a `$\rho$-pole'
contribution.}  This yields the results given in the second
and third columns of Table 2 for two parameterizations of the Omn\`es function,
the Pad\'{e} form, Eq. \ref{DPade}, and the numerical parameterization
given by Gasser
respectively.  Interestingly, $\beta_1$
develops a rather large imaginary part as can be seen in the second and
third column of Table 2.

The advantage of this approach is twofold. First we use the experimentally
available information on coefficients $\alpha_1$, $\beta_1$ -- the
shortcomings of a too low $K\rightarrow 3\pi$ amplitude from soft pion theorems
are thus avoided. Secondly, it provides us with a representation of real and
imaginary part of the decay amplitude outside the physical region. This is
exactly what is needed in order to make Im $C(s)$, Eq. \ref{ImCimproved},
approximately
real. If instead we would use the experimental expansion of $A_{+-0}$,
Eq. \ref{DPexpansion},
as an extrapolation, Im $C(s)$ would develop an unacceptable large imaginary
part at rather low energies, i.e. for $\sqrt{s} \geq 450 {\rm MeV}$. Since our
solution to the $K\rightarrow 3\pi$ amplitude is subject to several
approximations, we cannot hope that the imaginary part of Im $C(s)$ cancels
completely. However, it cancels to a large extent, i.e. it never reaches
ten percent of the real part in the region from two pion threshold up to
$\sqrt{s} \approx 600$ MeV.

We are now ready to calculate the function $C(s)$ using the improved
representation of its imaginary part. We used two parameterizations of the
Omn\`es function $D_0^{-1}$, the Pad\'{e} solution Eq. \ref{DPade} and
the numerical
representation given by Gasser. In Figure 4 a) the
improved imaginary
part Im $C(s)$ is compared to the lowest order approximation, and in
Figure 4 b)
the corresponding real parts calculated from the dispersion integral are
shown. In the dispersive approach, both imaginary and real part of $C(s)$
are rather enhanced, already just above the two pion threshold. Calculating
$BR(K_L\rightarrow \pi^0 \gamma\gamma)$ with these inputs we find a net
enhancement over the one-loop ChPT result by a factor
$\approx 1.9$.  The results
are summarized in Table 3  where the branching ratio is given
for several scenarios.

\begin{table}
\begin{center}
\begin{tabular}{|l|l|l|}
\hline
{\em input}  &  $BR [10^{-6}]$  &  $BR^{\rm abs} [10^{-6}]$\\
\hline
exp  &  $1.7\pm 0.3$  &  --- \\
\hline
ChPT (pion-loop)  &  0.59  &  0.39  \\
\hline
Pad\'{e}-charged  & 0.93  &  0.68 \\
\hline
Pad\'{e}-charged+neutral  & 1.12  &  0.84 \\
\hline
Gasser-charged  &  0.93   & 0.68   \\
\hline
Gasser-charged+neutral & 1.08  & 0.81 \\
\hline
\end{tabular}
\caption{Shown are the calculated branching ratios for
$K_L\rightarrow\pi^0\gamma\gamma$ under various input assumptions.
The second column indicates the contribution to the branching ratio
from the absorptive part of the amplitude.}
\end{center}
\end{table}

The observed enhancement is clearly seen to come from
two effects:\footnote{It should be noted that our result with inclusion of
charged intermediate state only is consistent with that previously calculated
by Ko and Rosner \cite{KR} using a simple one-loop approximation. We do not
understand the discrepancy with the dispersive calculation of Truong
\cite{Tru3} who also includes only the charged intermediate state and finds
a branching ratio of $1.3 \times 10^{-6}$.}
\begin{itemize}
\item[ i)]
The use of a corrected $K\rightarrow 3\pi$ amplitude in agreement with data
gives an enhancement factor of $\approx 1.6$ in the branching ratio. This has
been noted earlier \cite{KR,Eck}; in a very simple approach one could just
scale $G_8$ in Eq. \ref{K3piCA} to reproduce the experimental
$K\rightarrow 3\pi$ amplitude,
leading to a similar enhancement factor for
$BR(K_L\rightarrow \pi^0 \gamma\gamma)$. However, a consistent calculation
clearly should explain all relevant processes, $K\rightarrow 2\pi$,
$K\rightarrow 3\pi$, $\gamma\gamma\rightarrow \pi\pi$, $K_S\rightarrow
\gamma\gamma$ and $K_L\rightarrow \pi^0\gamma\gamma$ by the same method. We
have explained in detail above how this can be achieved.
\item[ii)]
The inclusion of the neutral two pion intermediate state in the unitarity
relation brings an additional enhancement factor of $\approx 1.2$.  As
mentioned
before, this contribution is missed in all approaches where the $\pi\pi
\rightarrow \gamma\gamma$ vertex is treated only at the Born level. Although
this effect is moderate, it goes in the right direction.  Moreover, it is
critical to include the $\pi^0\pi^0$ intermediate state in order to properly
implement the constraints of unitarity.
\end{itemize}

Finally we note that using a twice subtracted dispersion relation, the
contributions from the high energy region to the dispersive integral are
very much suppressed. Cutting the integral at $\sqrt{s}=600 {\rm MeV}$ instead
of 1 GeV changes the branching ratio by less than one percent.  Also,
the calculated spectrum in the invariant mass of the two photons is
plotted in Figure 3.  It deviates insignificantly from the spectrum obtained
in one loop Chiral perturbation theory.  If only the charged intermediate state
is included, the maximum of ${1\over \Gamma}{d\Gamma\over dz}$ is shifted
toward higher $q^2$ values, contrary to the experimental trend.  Inclusion
of the neutral intermediate state in the unitarity relation restores the
maximum
to its original location.

\section{Conclusions}

The process $K_L\rightarrow\pi^0\gamma\gamma$ has traditionally been a
difficult one to understand within the context of chiral perturbation theory.
Indeed, from other successes one might have expected the one-loop chiral
prediction to be accurate to ${\cal O}$(20\%) or so, whereas in fact the
predicted branching ratio is nearly a factor of three too small.  Previous
work in this area has attempted to explain this discrepancy in terms of
vector meson pole contributions and/or in terms of higher loop final
state interaction effects.
We have here noted, however, that previous dispersion-based final
state interaction calculations have included only the intermediate $\pi^+
\pi^-$ intermediate state, omitting its potentially important $\pi^0\pi^0$
analog.  Above we have given a mutually consistent analysis of both
$K_L\rightarrow3\pi$ {\it and} $K_L\rightarrow\pi^0\gamma\gamma$ processes
using dispersion relations and have shown that by including this previously
neglected $\pi^0\pi^0$ intermediate state piece the branching ratio for
$K_L\rightarrow \pi^0\gamma\gamma$ is significantly enhanced, although it
remains too low to fully explain the data.
This should by no means be considered to
be complete analysis---indeed many effects such as I=1,2 $\pi\pi$ scattering
effects as well as higher order contributions to the $K_L\rightarrow 3\pi$
process have been neglected.  Nevertheless, we believe that this calculation
opens up interesting questions for future study about the importance of
effects beyond the one loop approximation in chiral perturbation theory.

\vspace{1 cm}
\noindent
{\large \bf Acknowledgements}

We would like to thank J.F. Donoghue for helpful discussions. J.K. acknowledges
generous financial support from Schweizerischer Nationalfonds.

\pagebreak

\noindent
{\large \bf Figure captions}
\begin{itemize}
\item[Fig. 1: ]
Indicated are experimental data points for $\gamma\gamma\rightarrow\pi\pi$
compared to the one-loop chiral perturbative prediction (dashed) and dispersive
calculations using Pad\'e (dotted) and Gasser (solid) $D_0^{-1}(s)$ functions.
\item[Fig. 2: ]
Shown are the results of one-loop chiral perturbation theory (dashed) and
dispersive analysis scenario d) (solid) for Im $B(s)$ (a) and Re $B(s)$
(b) respectively.
\item[Fig. 3: ]
Normalized spectra $1/\Gamma d\Gamma/dz$ in the invariant mass of the two
photons ($z=q^2/M_{K^0}^2$). Plotted are chiral perturbation theory (dashed)
and dispersive analysis using Gasser $D_0^{-1}(s)$ function (solid).
\item[Fig. 4: ]
Im $C(s)$ (a) and Re $C(s)$ (b) in the physical decay region. Shown are
the results of one-loop chiral perturbation theory (dashed) and dispersive
analysis using Gasser $D_0^{-1}(s)$ function (solid). Also shown is the
result of the dispersive analysis including the charged intermediate state
only (dot-dashed).
\end{itemize}

\end{document}